\documentclass[useAMS,usenatbib]{mn2e}

\usepackage[american]{babel}
\usepackage{mathrsfs,eucal,graphicx,color,amsmath,amssymb,amsfonts}

\newcommand{\de}{\mathrm d}

\newcommand{\om}{\Omega_m}
\newcommand{\ob}{\Omega_b}
\newcommand{\odm}{\Omega_\mathrm{DM}}
\newcommand{\ode}{\Omega_\mathrm{DE}}

\newcommand{\ho}{{H_0}}

\newcommand{\lcdm}{$\Lambda$CDM}

\newcommand{\fom}{\mathrm{FoM}}

\title[Impact of Redshift Information on Cosmological Applications with Next-Generation Radio Surveys]{Impact of Redshift Information on Cosmological Applications with Next-Generation Radio Surveys}
\author[S. Camera et al.]{Stefano Camera,$^1$\thanks{E-mail: stefano.camera@ist.utl.pt (SC); mgrsantos@ist.utl.pt (MGS); david.bacon@port.ac.uk (DJB).} M\'ario G. Santos,$^{1\star}$ David J. Bacon,$^{2\star}$ Matt J. Jarvis,$^{3,4}$\newauthor Kim McAlpine,$^5$ Ray P. Norris,$^{6}$ Alvise Raccanelli$^{7,8}$ and Huub R\"ottgering$^9$\\$^1$CENTRA, Instituto Superior T\'ecnico, Universidade T\'ecnica de Lisboa, Av. Rovisco Pais 1, 1049-001 Lisboa, Portugal\\$^2$Institute of Cosmology \& Gravitation, University of Portsmouth, Dennis Sciama building, P01 3FX Portsmouth, UK\\$^3$Centre for Astrophysics Research, STRI, University of Hertfordshire, Hatfield, AL10 9AB, UK\\$^4$Department of Physics, University of the Western Cape, Bellville 7535, South Africa\\$^5$Department of Physics and Electronics, Rhodes University, Grahamstown 6139, South Africa\\$^6$CSIRO Astronomy \& Space Science, Australia Telescope National Facility, P.O. box 76, NSW 1710 Epping, Australia\\$^7$Jet Propulsion Laboratory, California 
Institute of Technology, Pasadena, CA 91109, USA\\$^8$California Institute of Technology, Pasadena, CA 91125, USA\\$^9$Leiden Observatory, University of Leiden, Leiden NL-2300 RA, The Netherlands}

\begin{document}

\date{Accepted 0000 --- 00. Received 0000 --- 00; in original form 0000 --- 00}

\pagerange{\pageref{firstpage}--\pageref{lastpage}} \pubyear{2012}

\maketitle

\label{firstpage}

\begin{abstract}
In this paper, we explore how the forthcoming generation of large-scale radio continuum surveys, with the inclusion of some degree of redshift information, can constrain cosmological parameters. By cross-matching these radio surveys with shallow optical to near-infrared surveys, we can essentially separate the source distribution into a low- and a high-redshift sample, thus providing a constraint on the evolution of cosmological parameters such as those related to dark energy. We examine two radio surveys, the Evolutionary Map of the Universe (EMU) and the Westerbork Observations of the Deep APERTIF Northern sky (WODAN). A crucial advantage is their combined potential to provide a deep, full-sky survey. The surveys used for the cross-identifications are SkyMapper and SDSS, for the southern and northern skies, respectively. We concentrate on the galaxy clustering angular power spectrum as our benchmark observable, and find that the possibility of including such low redshift information yields major improvements in the determination of cosmological parameters. With this approach, and provided a good knowledge of the galaxy bias evolution, we are able to put strict constraints on the dark energy parameters, i.e. $w_0=-0.9\pm0.041$ and $w_a=-0.24\pm0.13$, with type Ia supernov\ae\ and CMB priors (with a one-parameter bias in this case); this corresponds to a Figure of Merit (FoM) $>600$, which is twice better than what is obtained by using only the cross-identified sources and greater than four time better than the case without any redshift information at all.
\end{abstract}


\section{Introduction}
A new era for precision cosmology with radio continuum surveys is close to becoming a reality. Several experiments across the world are currently beginning operations, such as the LOw Frequency ARray\footnote{http://www.lofar.org} (LOFAR) and APERTIF\footnote{http://www.astron.nl/general/apertif/apertif} at the Westerbork Synthesis Radio Telescope (WSRT), in the northern hemisphere, and the projects for the Australian\footnote{http://www.atnf.csiro.au/projects/askap/} and African\footnote{http://www.ska.ac.za/index.php} SKA Pathfinders, in the southern hemisphere.

Recently, \citet{Raccanelli:2011pu} have shown promising forecasts for constraining cosmological parameters using some of the large-scale radio continuum surveys planned for these forthcoming telescopes. In this paper, we investigate the usefulness of including redshift information, for the subsample of radio continuum sources which are successfully cross-identified with optical surveys containing photometric or spectroscopic redshifts. We examine the impact of this partial redshift knowledge on cosmological constraints with forthcoming radio surveys. The underlying effect of this multi-wavelength information is reasonably simple to understand: shallow optical surveys should be able to identify most of the low redshift galaxies from these large radio surveys, allowing us to ``extract'' the high redshift tail expected for the radio distribution of sources.

Although the detailed analysis is more involved, basically the combination of these two ``macro'' redshift bins will give us a powerful constraint on the redshift evolution of cosmological models as an approximate alternative to demanding redshift surveys. For the purpose of this work and in order to clarify the improvements achieved with this redshift information, we will focus on the radio galaxy two-point correlation function as the cosmological observable; there are several other available probes such as the integrated Sachs-Wolfe effect and cosmic magnification---as discussed in \citet{Raccanelli:2011pu}---which will merit further investigation in the context of specific modified matter and gravity theories \citep{Camera:radio-mg}.

The paper is structured as follows. We briefly present the EMU and WODAN surveys and their main characteristics in \S~\ref{sec:surveys}, while the surveys used for  cross-identifications, providing redshift information, are described in \S~\ref{sec:XID}. \S~\ref{sec:pipe} outlines the data analysis procedure which we will consider for obtaining cosmological constraints. In \S~\ref{sec:spectra}, we give the most important formul\ae\ for the galaxy clustering angular power spectrum and its statistics. In \S~\ref{sec:models}, we introduce the fiducial dark energy (DE) model used in our analyses and in \S~\ref{sec:method} we describe our method. In \S~\ref{sec:results} we present the analysis of the combined radio and optical surveys for constraining the DE equation of state. First, we present the cosmological constraints available when there is no redshift information at all; secondly, we show what would be obtained in the idealised case of full knowledge of every radio-source redshift. Finally, we describe the 
realistic scenario expected to be available in the recent future, where we cross-identify as many EMU/WODAN galaxies as possible with optical redshift surveys, in particular SkyMapper and SDSS. For this last case, we scrutinise a simple one-parameter model for the radio-galaxy bias, as well as a two-parameter model which accounts for a further redshift dependence. Conclusions are drawn in \S~\ref{sec:conclusions}.

\section{Forthcoming Radio Surveys}\label{sec:surveys}
\begin{figure}
\centering
\includegraphics[width=0.5\textwidth]{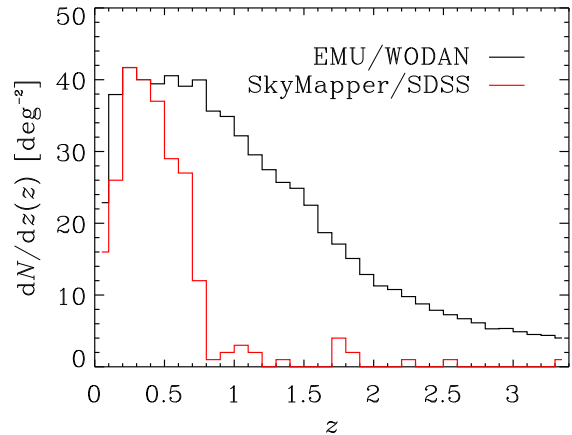}
\caption{Redshift distribution of sources $\de N/\de z(z)$ for the EMU/WODAN surveys. The area is normalised to the mean galaxy number density. The red histogram shows the distribution of radio objects which have optical counterparts in SkyMapper and SDSS.}\label{fig:dNdz}
\end{figure}

Here, we present the two survey designs we use in our calculations. A summary of their properties is given in Table~\ref{tab:EMUWODAN}, where $N_g$ represents the total number of detected radio galaxies at $10\sigma$ and $z_m$ the median redshift (see \citealt{Raccanelli:2011pu} for details). It is worth noticing that these two surveys have the same redshift distribution of sources, shown in Fig.~\ref{fig:dNdz} (black histogram). Furthermore, they will cover the entire sky, if their data are combined. This is one of the major strengths of the present analysis, providing a homogeneous all-sky catalogue, since the two surveys have basically the same sensitivity, galaxy number density and redshift distribution.
\begin{table}
\caption{Specifics of the EMU and WODAN surveys, where $N_g$ is the total number of detected sources at $10\sigma$, and $z_m$ the median redshift.}\label{tab:EMUWODAN}
\centering
\begin{tabular}{cccccc}
Survey & Area & Frequency & Sensitivity & $N_g$ & $z_m$ \\
\hline
EMU & $3\pi$ & $1400\,\mathrm{MHz}$ & $10\,\mu\mathrm{Jy}$ & $2.2\cdot10^7$ & $1.1$ \\
WODAN & $\pi$ & $1400\,\mathrm{MHz}$ & $10\,\mu\mathrm{Jy}$ & $7.3\cdot10^6$ & $1.1$
\end{tabular}
\end{table}

\subsection{EMU}\label{ssec:emu}
EMU (Evolutionary Map of the Universe, \citealt{Norris:2011ai}) is an all-sky continuum survey planned for the new Australian SKA Pathfinder (ASKAP, \citealt{2008ExA....22..151J}) telescope under construction on the Australian candidate SKA site in Western Australia. The primary goal of EMU is to make a deep $10\,\mu\mathrm{Jy/beam}$ radio continuum survey of the entire Southern Sky, extending as far North as $+30^\circ$ at a resolution of $\sim 10$\,arcsec; it will also have sensitivity to extended structures. The EMU survey is expected to begin in $2013$.

\subsection{WODAN}\label{ssec:wodan}
WODAN (Westerbork Observations of the Deep Apertif Northern sky survey) is planned to chart the entire northern sky above $+30^\circ$ down to a proposed rms flux density at $1.4\,\mathrm{GHz}$ of $10\,\mu\mathrm{Jy}$ per beam with a resolution of $\sim 15$\,arcsec \citep{Rottgering:2011jq}. It will be able to do this because of the new phased array feeds (APERTIF) being put on the Westerbork Synthesis Radio Telescope (WSRT, \citealt{Oosterloo:2010wz}). The current schedule for the commencement of APERTIF surveys is 2013.

LOFAR is another experiment which merits comment. This is a multi-national telescope with stations spanning Europe; the core of LOFAR is situated in the north-east of the Netherlands, with stations on longer baselines both within the Netherlands and across Germany, UK, France and Sweden. Other stations may also be added throughout the rest of Europe in the coming years. LOFAR large-area continuum surveys probe to similar depth and source densities as EMU and WODAN \citep{Rottgering:2003jh,Rottgering:2011jq,vanHaarlem:2012}, and can therefore be used on their own or in conjunction with EMU and WODAN for cosmological studies. This is because the lower frequency of LOFAR makes it sensitive to different source populations and provides spectral information. However, for the rest of this paper we concentrate on only the EMU and WODAN surveys, noting that LOFAR could also be used in the northern hemisphere.

\section{Surveys for Cross-Identifications}\label{sec:XID}
In this section, we present the surveys we shall use to cross-identify the EMU/WODAN sources. Our goal is to obtain any sort of redshift information for some of the radio sources that will be detected by the EMU and WODAN surveys using surveys available around the same time. We shall mainly focus on current and forthcoming large-scale surveys which have multi-filter data across the optical and near-infrared to facilitate reasonably accurate photometric redshifts, either through pure template fitting methods \citep[e.g.][]{Benitez:1998br,Bolzonella:2000js}, template fitting methods combined with known spectroscopic redshifts \citep[e.g.][]{Feldmann:2006wg, Ilbert:2006dp,Salvato:2011bz} or with empirical methods, such as artificial neural networks \citep[e.g.][]{Firth:2002yz, Collister:2003cz} or Gaussian Processes \citep[e.g.][]{Way:2006pz, 2010MNRAS.405..987B}.

The most important surveys to consider for these radio surveys are therefore SkyMapper \citep{Keller:2007xt} and SDSS \citep{Eisenstein:2011sa} at optical wavelengths, combined with the UKIRT Infrared Deep Sky Survey Large Area Survey \citep[UKIDSS-LAS; ][]{Lawrence:2006de} and the proposed UKIRT Hemisphere Survey at northern latitudes, along the VISTA Hemisphere Survey at Southern Latitudes. We also note that the Wide-field Infrared Survey Explorer \citep[WISE; ]{Wright:2010qw} could also play a r\^ole in helping with photometric redshifts over both hemispheres. Further in the future, there are deeper planned surveys observing large areas, such as those using  the Large Synoptic Survey Telescope \citep[LSST; ][]{Ivezic:2008fe} and Euclid\footnote{http://www.euclid-ec.org} \citep{EditorialTeam:2011mu}. We do not consider the slightly smaller area surveys such as the Dark Energy Survey \citep[DES;][]{Abbott:2005bi} in this paper, but note that they could provide important additional information by 
constraining redshifts to fainter magnitudes and higher redshifts.

In this paper, we focus on SkyMapper and SDSS. The former is a $20,000\,\mathrm{deg}^2$ optical survey of the southern sky, whilst the latter will cover $12,000\,\mathrm{deg}^2$ in the northern sky. For both surveys, we adopt a conservative magnitude limit of $r<22\,\mathrm{mag}$ which should ensure that the redshifts can be measured to reasonable accuracy and certainly to the level required for our investigations here. SkyMapper should provide photometric redshifts whilst SDSS should be able to also provide spectroscopic redshifts---at least for low $z$'s.

To assess the number of radio sources in both WODAN and EMU that will have an optical counterpart to $r\sim 22$, we use the recent results of \citet{McAlpine:2012cu}, who analysed how many reliable cross-identifications to radio sources can be found as a function of resolution and depth of the radio data coupled with the depth of the optical and near-infrared data, using the VISTA Deep Extragalactic Observaitions (VIDEO) Survey \citep{Jarvis:2012zy}. In Fig.~\ref{fig:dNdz} we show the expected redshift distribution of radio sources detected at $>10\sigma$ in EMU and WODAN (black) along with the expected number of optical counterparts at $r<22$ from SDSS and SkyMapper (red).

\section{Strategy for Redshift Binning}\label{sec:pipe}
In this section, we describe a realistic scenario for an analysis leading to cosmological constraints including redshift information. First, we note that the redshift distribution shown in Fig.~\ref{fig:dNdz} should be known with reasonable accuracy for forthcoming radio surveys. This is because there are radio surveys such as the Australia Telescope Large Area Survey\citep[ATLAS; ][]{Norris:2006er} and radio surveys covering the UKIDSS-Ultra Deep Survey \citep{Simpson:2006if} and the COSMOS field \citep{Schinnerer:2010jz}, covering a few square degrees to the same depth as EMU and WODAN, and for which deep spectroscopic data is becoming available. We shall consider an analysis where a fraction of the EMU/WODAN galaxies are cross-identified with SkyMapper and/or SDSS. Hence, for those sources we shall be able to perform an actual redshift binning.

We assume that spectroscopic redshifts have a negligible uncertainty, and that they will be available for the radio sources using the SDSS up to $z\simeq0.3$. For those sources without spectroscopic redshifts, namely radio sources with  identifications in SkyMapper and those without spectroscopic redshifts in SDSS (i.e. with $z> 0.3$) we shall rely on photometric redshifts and assume a photometric-redshift scatter $\sigma^\mathrm{ph}_z=0.05(1+z)$. The binning we use is summarised in Table~\ref{tab:z-errors}. Note that this does not refer to all the sources that will be detected by these optical surveys, but only to those that are cross-matched to EMU and WODAN. In particular, radio AGNs detected in the radio occupy massive galaxies, which should correspond to the SDSS LRGs (Luminous Red Galaxies), so in principle we could have spectroscopic information from SDSS to higher redshifts (e.g. up to $z\sim 0.5$ instead of $z\lesssim 0.3$ that we assume here).
\begin{table}
\caption{Adopted bins for cross-identifications with SkyMapper and SDSS (equal size bins for each range). The assumed scatter for photo-$z$ bins is $0.05(1+z)$ for photo-$z$.}\label{tab:z-errors}
\centering
\begin{tabular}{ccc}
& SkyMapper & SDSS \\
\hline
Spectroscopic & $-$ & $3$ bins $(0.0<z<0.3)$ \\
Photometric & $5$ bins $(0.0<z<0.8)$ & $3$ bins $(0.3<z\lesssim0.8)$
\end{tabular}
\end{table}

On the other hand, one of the most promising properties of EMU and WODAN is the high-redshift tail of their source distribution (see Fig.~\ref{fig:dNdz}), which is composed of galaxies difficult to cross-identify with SkyMapper or SDSS due to their faint nature. Nevertheless, these can still bring important additional information to the analysis. We will assess the impact of these objects by making calculations for both a cross-identified catalogue, and a catalogue whose source distribution is given by the difference between the EMU/WODAN source distribution and that of the survey used for the cross-identifications. In addition, we note that the sky coverage of the optical surveys will not overlap exactly with EMU or WODAN. Therefore, we can also make calculations for a third catalogue which follows the EMU/WODAN source distribution and covers all the sky not surveyed by the optical surveys.

To summarise, the analysis for this realistic-binning strategy will combine three ``effective surveys,'' whose $\de N/\de z(z)$ are defined as follows:
\begin{itemize}
\item[\textit{i)}] all the EMU/WODAN galaxies cross-identified by SkyMapper/SDSS, appropriately binned in redshift (see Table~\ref{tab:z-errors});\\
\item[\textit{ii)}] all the EMU/WODAN galaxies which are in the same patch of sky as SkyMapper/SDSS but which are not cross-identified;\\
\item[\textit{iii)}] the part of the EMU/WODAN survey probing the sky left uncovered by SkyMapper/SDSS.
\end{itemize}

\section{Galaxy Angular Power Spectrum}\label{sec:spectra}
We now introduce the main cosmological observable considered in this work. For our radio surveys, let us consider the 2-point angular correlation function $\xi^g$ of radio sources. It can be seen as the excess probability $\delta_g$ of finding two radio sources at a certain physical separation $s=|\mathbf x-\mathbf y|$ one from each other \citep{1980lssu.book.....P}, viz.
\begin{equation}
\langle\delta_g(\mathbf x)\delta_g(\mathbf y)\rangle\equiv\delta_D(s)\xi^g(s);
\end{equation}
here, $\delta_D$ is Dirac's delta function. By performing the Fourier transform of such an observable we obtain the radio source over-density power spectrum $P^g(k,z)$, where $\mathbf k$ is the physical wavenumber, with $k=|\mathbf k|$. This is related to the underlying total-matter power spectrum $P^\delta(k,z)$ of the matter density contrast $\delta\equiv\delta\rho/\rho$ by the radio-source bias function $b_g$, which we discuss in greater detail in \S~\ref{ssec:bias}. Thus, we have $P^g(k,z)={b_g}^2P^\delta(k,z)$.

However, when we have no radial information, we instead deal with projected quantities. Thus, the angular power spectrum $C^g(\ell)$ of the radio source fluctuation---where $\ell$ is the angular wavenumber---then reads
\begin{equation}
C^g(\ell)=4\pi\int\!\!\frac{\de k}{k}\,\left[W^g(\ell,k)\right]^2P^g(k,z=0),
\end{equation}
with $W^g(\ell,k)$ a proper line-of-site weight function. A widely used simplification is given by the so-called Limber's approximation \citep{Kaiser:1991qi}, where $\ell=k\chi$; here, $\chi(z)$ is the radial comoving distance, which is related to the Hubble expansion rate $H(z)$ via $\de\chi=\de z/H(z)$. Limber's approximation is valid when $\ell\gg1$, but it has been shown that the convergence is already good for $\ell\gtrsim10$ \citep[e.g.][]{Hu:2000ee}. Therefore, it is a suitable approximation, since for larger angular scales the cosmic variance uncertainty is dominant. In this limit we have
\begin{equation}
C^g(\ell)=\int\!\!\de\chi\,\left[\frac{W^g(\chi)}{\chi}\right]^2P^\delta\!\left(\frac{\ell}{\chi},\chi\right),\label{eq:C^g_l}
\end{equation}
with $W^g(\chi)$ defined through
\begin{equation}
W^g[\chi(z)]=H(z)b_g(z)\frac{\de N}{\de z}(z).\label{eq:W^g}
\end{equation}

\subsection{Radio-Galaxy Bias}\label{ssec:bias}
\begin{figure}
\centering
\includegraphics[width=0.5\textwidth]{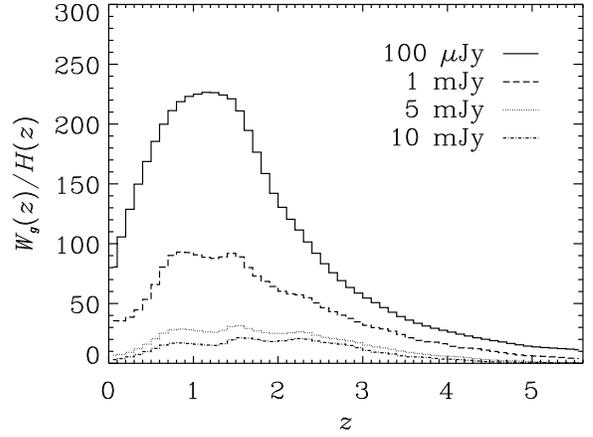}
\caption{The bias multiplied by the redshift distribution of the radio sources as a function of redshift for the EMU and WODAN surveys (solid line). Also shown are the product of the bias and redshift distribution for surveys at 1\,mJy, 5\,mJy and 10\,mJy to illustrate the advance that is expected from the new generation of radio surveys compared to surveys such as the NVSS. For the functional forms of the bias of each source population, we refer to \citet[][Fig.~3; see also \citealt{Wilman:2008ew}]{Raccanelli:2011pu}.}\label{fig:bias}
\end{figure}

The radio-source bias is worth a deeper digression. In this paper, we adopt the approach of \citet{Raccanelli:2011pu}. They used the simulations of \citet{Wilman:2008ew} (see also \citealt{Wilman:2010sp}) for the SKA continuum survey. By generating source catalogues with the $S$-cube database\footnote{http://s-cubed.physics.ox.ac.uk} corresponding to the radio flux-density limits of the proposed EMU and WODAN surveys, they obtained an estimate of the bias for each source population used here. Note that the source distribution and bias of \citet{Wilman:2008ew} are compatible with several observational results such as the NVSS survey \citep{1998AJ....115.1693C}. In Fig.~\ref{fig:bias}, we show the product of the bias and the redshift distribution of the sources, i.e. $W_g(z)/H(z)$; we also illustrate the products of the bias and redshift distribution for surveys at 1\,mJy, 5\,mJy and 10\,mJy, to illustrate the advance that is expected from the new generation of radio surveys compared to surveys such as the NVSS.

In this paper, we assume a reasonable knowledge \textit{a priori} of the bias evolution, so that only a one- and a two-parameter model are used, as explained below.  The analysis of small sky patches with deep radio data and detailed redshift information from other multi-wavelength data should provide a reasonable measurement of the bias directly from the redshift space distortions without assumptions on the DE model, and thus allow for this more simple parameterisation (note that the bias is assumed scale independent).  Nevertheless, as constraints on the DE evolution become more stringent, a more precise procedure will be required, such as a joint fit to both the bias and DE model using the deep, small survey and the large, shallow radio survey combined.  Specifically, from correlation function of the faint radio sources from the deep survey, we shall have important information on the galaxy clustering as a function of redshift, whilst the data from shallow, large-scale surveys will provide information on the scale-dependent effects of DE.  Moreover, this could include other estimators such as the ISW and, in particular, the cosmic magnification, which is less sensitive to the bias issue.  Finally, other possible parameterisations of the bias within the theoretical framework of the halo model \citep[e.g.][for a review]{Cooray:2002dia} should be explored; they could allow us to reduce the number of parameters.  The current paper addresses the constraining capabilities of the next generation of radio surveys that will soon be available, when combined with other multi-wavelength data using the clustering estimator as a benchmark, and we hope that the results obtained open the window for further studies.

The effective bias $b_g(z)$ of Eq.~\eqref{eq:W^g} accounts for the different biases for each source population. The main results of this work (\S~\ref{ssec:realistic}) have been obtained by allowing for the amplitude of this effective bias to take any value, while maintaining the redshift evolution of \citet[][their Fig.~3]{Raccanelli:2011pu}. That is to say, we parameterise the effective bias as $A_bb_g(z)$, with a fiducial value $A_b=1$, and we then marginalise the whole Fisher matrix over such amplitude.

Nonetheless, we also study the effects of allowing for a greater uncertainty in the redshift evolution of the bias by introducing another parameter that accounts for an additional redshift dependence. Specifically, we make the substitution
\begin{equation}
A_bb_g(z)\to\left(A_b+B_bz\right)b_g(z),\label{eq:lin-bias}
\end{equation}
with $B_b=0$ as a fiducial value. We shall see that this will not qualitatively change our results---though the presence of one more free parameter necessarily broadens the constraints.

\section{Dynamical Dark Energy}\label{sec:models}
Here, we briefly outline the cosmological model tested in our analysis. Since the objective of this work is to test the effects of redshift information, we choose a dynamical DE model as our benchmark. We consider a flat Friedmann-Lema\^itre-Robertson-Walker Universe, filled with a perfect fluid of baryons, cold dark matter and DE. Their abundances in units of the critical density are $\ob$, $\odm$ and $\ode=1-\om$, respectively, and the total matter contribution is $\om\equiv\ob+\odm$. Therefore, the expansion history of the Universe $H\equiv\de\ln a/\de t$ is \citep[see e.g.][]{Bartelmann:2009te}
\begin{equation}
H(z)=\ho\sqrt{\om\left(1+z\right)^3+\ode f_\mathrm{DE}(z)},\label{eq:hubble}
\end{equation}
with $\ho=100\,h\,\mathrm{km/s/Mpc}$ the Hubble constant and $f_\mathrm{DE}(z)$ the (unknown) function governing the time evolution of DE.

For an exhaustive and recent review on DE in general, we suggest to the reader the comprehensive monograph of \citet{2010deto.book.....A}. For the purpose of this work, it is sufficient to say that, in principle, there is a time-dependent equation of state linking the DE pressure to its energy density, namely
\begin{equation}
p_\mathrm{DE}(a)=w_\mathrm{DE}(a)\rho_\mathrm{DE}(a),
\end{equation}
where $a=1/(1+z)$ is the scale factor. Thus, $f_\mathrm{DE}(z)$ in Eq.~\eqref{eq:hubble} is
\begin{equation}
f_\mathrm{DE}(z)=\exp\left[-3\int_1^a\!\!\de a'\frac{1+w_\mathrm{DE}(a')}{a'}\right].
\end{equation}
Hereafter, we shall use the well-known Chevallier-Polarski-Linder parameterisation \citep{Chevallier:2000qy,Linder:2002et}
\begin{equation}
w_\mathrm{DE}(a)=w_0+(1-a)w_a,
\end{equation}
which is a Taylor expansion around a cosmological constant (today). Indeed, the \lcdm\ limit is reached when $w_0\to-1$ and $w_a\to0$.

\begin{figure}
\centering
\includegraphics[width=0.5\textwidth,trim=20mm 0mm -10mm 0mm,clip]{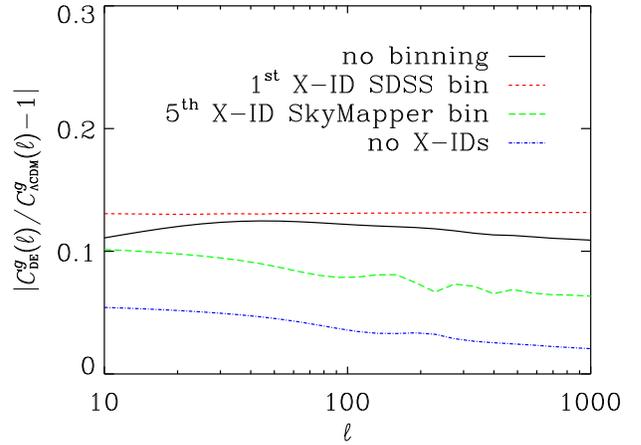}
\caption{Relative difference between DE and \lcdm\ galaxy clustering angular power spectra. The black (solid) curve refers to the unbinned EMU+WODAN case, the red (short-dashed) and the green (long-dashed) curves correspond---according to Table~\ref{tab:z-errors}---to the first SDSS and the last SkyMapper bin, respectively, and the blue (dot-dashed) line corresponds to the EMU/WODAN sources without cross-identifications (see Fig.~\ref{fig:dNdz}).}\label{fig:spectra}
\end{figure}

To highlight why including some redshift information is important for the aim of constraining DE with the radio source angular power spectrum, in Fig.~\ref{fig:spectra} we show the  $C^g(\ell)$ of Eq.~\eqref{eq:C^g_l} for different redshift bins. Ultimately, this means that the angular power spectra can in principle probe departures from standard \lcdm, if the signal-to-noise is sufficiently high. We plot the relative difference between the DE and \lcdm\ galaxy-clustering angular power spectra as a function of the angular scale $\ell$ for the different effective surveys introduced in \S~\ref{sec:pipe}. Specifically, the solid (black) curve refers to the standard, unbinned EMU+WODAN case, while the short-dashed (red) and the long-dashed (green) curves respectively correspond to the first SDSS and the last SkyMapper bin (see Fig.~\ref{fig:dNdz} and Table~\ref{tab:z-errors}); the dot-dashed (blue) line corresponds to the high-$z$ tail of the effective survey (\textit{ii}). In the next section we shall calculate 
the degree to which we can measure these departures from \lcdm\ given 
realistic source densities.

\section{Method and Fisher Matrix Formalism}\label{sec:method}
The expected errors on the model parameters can be estimated via the Fisher information matrix \citep{Fisher:1935,Jungman:1995bz,Tegmark:1996bz}. This has the advantage that different observational strategies can be analysed and this can be very valuable for experimental design. The Fisher matrix gives the best errors to expect, and is accurate if the likelihood surface near the peak is adequately approximated by a multivariate Gaussian. Hence, the Fisher matrix is simply the sum of the Fisher matrices of each $\ell$ mode,
\begin{equation}
\mathbfss F_{\alpha\beta}=\sum_{\ell=\ell_\mathrm{min}}^{\ell_\mathrm{max}}\frac{2\ell+1}{2}f_\mathrm{sky}\frac{\partial C^g(\ell)}{\partial\vartheta_\alpha}\left(\mathbfss C^{g,g}_\ell\right)^{-1}\frac{\partial C^g(\ell)}{\partial\vartheta_\beta},\label{eq:Fisher}
\end{equation}
where $f_\mathrm{sky}$ is the fraction of the sky covered by the survey under analysis and $\bvartheta=\{\om,\,\ob,\,w_0,\,w_a,\,h,\,n_s,\,\sigma_8,\,A_b\}$ is the parameter set, with $A_b$ the overall galaxy-bias amplitude (as introduced in \S~\ref{ssec:bias}). $\mathbfss C^{g,g}_\ell$ is the covariance for a given $\ell$ mode,
\begin{equation}
\mathbfss C^{g,g}_\ell=\left[C^g(\ell)+\frac{1}{N_g}\right]^2,\label{eq:Cov^g_l}
\end{equation}
where the second term is related to the Poisson noise, $N_g$ being the mean number density of sources per steradian.

We remind the reader that we are considering Limber's approximation, the sum in Eq.~\eqref{eq:Fisher} therefore starts at $\ell_\mathrm{min}=10$. To check the robustness of the results obtained under such assumption, we also performed the analysis with $\ell_\mathrm{min}=2$. However, we found that the enhancements thus yielded are only of a few percent. This is because, at such large scales, the cosmic variance heavily spoils the signal.  Hence, we find it reasonable to neglect $\ell<10$. Furthermore, \citet{Hu:2000ee} have shown that Limber's approximation leads to an overestimation of the signal for $\ell\lesssim10$, which means that the actual improvement in the analysis would be even smaller than what we have estimated. On the other hand, $\ell_\mathrm{max}=1000$ is kept fixed.

At this stage, the overall galaxy-bias amplitude $A_b$ represents a ``nuisance'' parameter, since this quantity is not one of the fundamental cosmological parameters that we want to constrain. This results in a general broadening of all the marginal errors, particularly of the power-spectrum normalisation $\sigma_8$, which is completely degenerate with the bias amplitude. Therefore, we marginalise over both these parameters \citep{Wang:2008zh}. To avoid any possible numerical instability in the marginalising procedure, we calculate the the Fisher matrix marginalised over $\sigma_8$ and $A_b$ as \citep{Albrecht:2009ct}
\begin{equation}
\mathbfss G=\mathbfss F^{\varphi\varphi}-\mathbfss F^{\varphi\psi}\mathbfss U\boldsymbol{\Lambda}^{-1}\mathbfss U^T\mathbfss F^{\varphi\psi},\label{eq:marg-F}
\end{equation}
where we define $\bvarphi=\{\om,\,\ob,\,w_0,\,w_a,\,h,\,n_s\}$ and $\bpsi=\{\sigma_8,\,A_b\}$; therefore, $\mathbfss F^{\varphi\varphi}$ is the block of the total Fisher matrix containing the parameters we want to constrain, whilst $\mathbfss F^{\psi\psi}$ is the nuisance-parameter Fisher sub-matrix. Here, $\boldsymbol{\Lambda}$ is the diagonal matrix whose elements are the eigenvalues of $\mathbfss F^{\psi\psi}$, whilst $\mathbfss U$ is the orthogonal matrix diagonalising $\mathbfss F^{\psi\psi}$. By using Eq.~\eqref{eq:marg-F}, our marginalising procedure is more stable, since degeneracies in $\mathbfss F^{\varphi\varphi}$ are properly propagated to $\mathbfss G$ with no instabilities, and we do not even worry about a possibly ill-conditioned $\mathbfss F^{\psi\psi}$ sub-matrix, since we check its stability on the fly by the diagonalisation.

Another important quantity that we calculate is the correlation between the parameter pair $\varphi_i$-$\varphi_j$, which is defined as\footnote{Note that indices $\alpha,\beta=1\ldots8$ refer to the complete set of parameters $\bvartheta$, whilst $i,j=1\ldots6$ label the sub-set $\bvarphi$.}
\begin{equation}
r_{\varphi_i-\varphi_j}=\frac{\left(\mathbfss G^{-1}\right)_{ij}}{\sqrt{\left(\mathbfss G^{-1}\right)_{ii}\left(\mathbfss G^{-1}\right)_{jj}}}.\label{eq:correlation}
\end{equation}
This quantity tells us whether the two parameters are completely uncorrelated (if it is zero) or completely degenerate (if it is $\pm1$), or have some intermediate level of correlation.

Finally, we calculate the DE Figure of Merit (FoM), as introduced by the Dark Energy Task Force \citep{Albrecht:2009ct}
\begin{equation}
\fom=\frac{1}{\sqrt{\det\left[\left(\mathbfss G^{-1}\right)_{w_0w_a}\right]}}\label{eq:fom};
\end{equation}
this is proportional to the inverse of the area encompassed by the ellipse representing the $68\%$ confidence level in the $(w_0,\,w_a)$-plane. Therefore, the tighter the constraints, the larger the $\fom$. We also combine all our results with the CMB Fisher matrix obtained for the Planck survey \citep{:2006uk} and the latest measurements of type Ia supernov\ae\ (SNeIa) from the Union2 sample \citep{Amanullah:2010vv}.

Since one of our main goals is to show the capability of our method of discriminating between the fiducial DE model and \lcdm, there is also another possible approach. There is a pivot redshift $z_p$ where the uncertainty on the equation-of-state parameter is minimised for a given data model. Hence, rather than using $w_0$, we can introduce
\begin{equation}
w_p=w_0-\frac{z_p}{1+z_p}w_a\label{eq:w_p}.
\end{equation}
It is also easy to demonstrate that $\fom=\left(\sigma_{w_p}\sigma_{w_a}\right)^{-1}$ holds \citep{Albrecht:2009ct}.

\section{Constraints on Dark Energy}\label{sec:results}
In order to assess the impact of using redshift information in our analysis, we start by considering a flat Universe in the matter-dominated era with matter fraction $\om=0.28$, baryonic contribution $\ob=0.045$, Hubble constant (in units of $100\,\mathrm{Mpc}^{-1}$) $h=0.7$, slope of the primordial power spectrum $n_s=0.96$ and rms mass fluctuations on the scale of $8\,h^{-1}\,\mathrm{Mpc}$ $\sigma_8=0.8$. The fiducial model we use is slightly away from the \lcdm\ point in the DE parameter space: $\{w_0,\,w_a\}=\{-0.9,\,-0.24\}$, as fitted by \citet{Zhao:2009ti} against a large number of available data. The transfer function for the scale dependence of matter density perturbations is calculated via the fitting formul\ae\ of \citet{Eisenstein:1997ik}, which take the baryonic features in the matter power spectrum into account.

We use the redshift distribution of sources of the EMU and WODAN surveys (Fig.~\ref{fig:dNdz}), as outlined in \S~\ref{ssec:emu}-\ref{ssec:wodan}. The median redshift of the survey is $z_m=1.1$. The total number of sources is $2.2\cdot10^7$, for EMU, and $7.3\cdot10^6$, for WODAN. The total sky coverage is $4\pi$, three quarters coming from EMU and one quarter from WODAN. As pointed out above, we should have access to this overall redshift distribution beforehand.

\subsection{No Redshift Information}
First, we present the constraints that can be put on the cosmological parameters if no redshift information is available. The marginal errors $\sigma_{\varphi_i}$ on the fiducial values $\mu_{\varphi_i}$ of the parameters $\varphi_i$, as well as the correlations between the parameters and $w_0$ or $w_a$, are shown in Table~\ref{tab:DE}. The numbers in brackets refer to results including the full Fisher matrices for Planck and Union2. These marginal errors are approximately of the same order of magnitude as the parameter fiducial value. However, when we add CMB and SNeIa information, the forecast significantly improves. Nevertheless, it is still impossible to discriminate between the \lcdm\ model and DE. This is because at the time of recombination the DE component was massively subdominant compared to matter and radiation, and the use of CMB is therefore not enough. Similarly, though they are a powerful probe for the late-time accelerated expansion of the cosmos, SNeIa are only sensitive to the background 
dynamics.
\begin{table*}
\caption{Cosmological constraints when assuming no redshift information for the EMU/WODAN sources. Marginal errors $\sigma_{\varphi_i}$ on the fiducial values $\mu_{\varphi_i}$ of the DE model parameters $\varphi_i$ and their correlation with $w_0$ and $w_a$, viz. $r_{\varphi_i-w_0}$ and $r_{\varphi_i-w_a}$; $\sigma_8$ and $b_g$ have been marginalised over. Numbers in brackets refer to the inclusion of Planck and Union2 Fisher matrices.}\label{tab:DE}
\centering
\begin{tabular}{ccrlrlrl}
\hline
$\varphi_i$ & $\mu_{\varphi_i}$ & \multicolumn{2}{c}{$\sigma_{\varphi_i}$} & \multicolumn{2}{c}{$r_{\varphi_i-w_0}$} & \multicolumn{2}{c}{$r_{\varphi_i-w_a}$} \\
\hline
$\om$ & $ 0.28 $ & $ 0.18 $ & $ (0.00080) $ & $ 0.78 $ & $ (-0.335) $ & $ -0.81 $ & $ (0.20) $ \\
$\ob$ & $ 0.045 $ & $ 0.055 $ & $ (0.0028) $ & $ 0.535 $ & $ (-0.62) $ & $ -0.555 $ & $ (0.63) $ \\
$w_0$ & $ -0.9 $ & $ 2.47 $ & $ (0.11) $ & $ 1 $ & $ (1) $ & $ -0.99 $ & $ (-0.985) $ \\
$w_a$ & $ -0.24 $ & $ 9.60 $ & $ (0.40) $ & $ -0.99 $ & $ (-0.985) $ & $ 1 $ & $ (1) $ \\
$h$ & $ 0.7 $ & $ 0.19 $ & $ (0.00036) $ & $ 0.026 $ & $ (0.62) $ & $ -0.056 $ & $ (-0.62) $ \\
$n_s$ & $ 0.96 $ & $ 0.14 $ & $ (0.0036) $ & $ -0.56 $ & $ (-0.06) $ & $ 0.62 $ & $ (0.19) $ \\
\hline
\end{tabular}
\end{table*}

A possible solution to this problem is to include in the analysis information from other cosmological probes, as has been done by \citet{Raccanelli:2011pu}. Alternatively, we can investigate how much information is hidden in the redshift distribution of the radio sources detected by EMU and WODAN.

\subsection{Full Redshift Information}
Before showing the results obtained with the realistic strategy outlined in \S~\ref{sec:pipe}, we briefly discuss an extreme idealistic case for comparison. Here, we assume we perfectly know the redshifts of all the EMU and WODAN sources. Then, if we used the same bins of width $0.1$ as shown in Fig.~\ref{fig:dNdz}, we would have at our disposal $33$ spectra combined into a tomographic matrix $C_{ij}^g(\ell)$. Each element of the matrix is a galaxy angular power spectrum as in Eq.~\eqref{eq:C^g_l}, with the fraction of sources contained in the corresponding bin. It is worth noting that such a tomographic matrix is diagonal, since any cross-correlation between different bins vanishes. In other words, sources in the $i$-th bin do not overlap in redshift with sources in the $j$-th bin.

The marginal errors $\sigma_{\varphi_i}$ on the fiducial values $\mu_{\varphi_i}$ of the parameters $\varphi_i$, as well as the correlations between the parameters and $w_0$ or $w_a$, are shown in Table~\ref{tab:DE-spectro}. The numbers in brackets refer to the results including the full Fisher matrices for Planck and Union2.
\begin{table*}
\caption{The same as Table~\ref{tab:DE} for the unrealistic case where we have precise redshift information from all the sources of the survey.}\label{tab:DE-spectro}
\centering
\begin{tabular}{ccrlrlrl}
\hline
$\varphi_i$ & $\mu_{\varphi_i}$ & \multicolumn{2}{c}{$\sigma_{\varphi_i}$} & \multicolumn{2}{c}{$r_{\varphi_i-w_0}$} & \multicolumn{2}{c}{$r_{\varphi_i-w_a}$} \\
\hline
$\om$ & $ 0.28 $ & $ 0.00425 $ & $ (0.00034) $ & $ 0.48 $ & $ (-0.27) $ & $ -0.68 $ & $ (0.059) $ \\
$\ob$ & $ 0.045 $ & $ 0.0013 $ & $ (0.00023) $ & $ 0.43 $ & $ (-0.0076) $ & $ -0.61 $ & $ (0.020) $ \\
$w_0$ & $ -0.9 $ & $ 0.022 $ & $ (0.012) $ & $ 1 $ & $ (1) $ & $ -0.92 $ & $ (-0.89) $ \\
$w_a$ & $ -0.24 $ & $ 0.058 $ & $ (0.033) $ & $ -0.92 $ & $ (-0.89) $ & $ 1 $ & $ (1) $ \\
$h$ & $ 0.7 $ & $ 0.0032 $ & $ (0.000029) $ & $ -0.044 $ & $ (0.47) $ & $ -0.23 $ & $ (-0.145) $ \\
$n_s$ & $ 0.96 $ & $ 0.0057 $ & $ (0.0021) $ & $ -0.18 $ & $ (0.32) $ & $ 0.46 $ & $ (-0.053) $ \\
\hline
\end{tabular}
\end{table*}

In this extreme case, the results are unsurprisingly excellent. The possibility of binning the EMU and WODAN sources allows us to track the evolution of DE up to high redshifts, thanks to the long tail of the distribution of the radio sources (see Fig.~\ref{fig:dNdz}). Indeed, even though DE is a subdominant species compared to matter until very late times, it is crucial to study its evolution, in order to constrain $w_a$. Moreover, the combination of both surveys yields a full-sky coverage, thus providing better statistics. As a result, the forecast marginal errors on the DE parameters are $\sigma_{w_0}=0.022$ and $\sigma_{w_a}=0.058$, which become $\sigma_{w_0}=0.012$ and $\sigma_{w_a}=0.033$ when combined with Planck and Union2. These constraints finally give $\fom$s of $1.9\cdot10^3$ and $5.4\cdot10^3$, respectively.

The correlation coefficients $r_{\varphi_i-w_0}$ and $r_{\varphi_i-w_a}$ are particularly useful to understand whether this present binning method is able of disentangling the parameter dependencies and lift some of their degeneracies. Indeed, if we look at the correlation coefficients in Table~\ref{tab:DE}, we find that some of them are near unity in absolute value. This means that these constraints are strongly degenerate with each other; this is because the closer to unity is the value of $r_{\varphi_i-\varphi_j}$, the more correlated are the two parameters $\varphi_i$ and $\varphi_j$. On the other hand, the correlations obtained in combination with Planck and Union2 are far better for all the parameters.

\subsection{Realistic Binning Strategy}\label{ssec:realistic}
Now, we present the realistic results that the procedure discussed in \S~\ref{sec:pipe} will yield. Since we are now dealing with the cross-identified sources, their redshifts (and the relative error we have to assume) are dependent on the optical survey used for the cross-identification, as summarised in Table~\ref{tab:z-errors}. The EMU radio sources will be identified by SkyMapper, which will provide photometric estimates for the source redshifts. Therefore, we divide the distribution $\de N/\de z(z)$ of Fig.~\ref{fig:dNdz} (red curve) into five bins up to $z=0.8$ and assume a scatter $\sigma^\mathrm{ph}_z=0.05(1+z)$. For WODAN, the situation is slightly more complicated. This is because SDSS will provide spectroscopic measurements for objects with $z\lesssim0.3$, whilst we will have to rely on photometric redshifts for the sources at higher redshift. Hence, in this case we subdivide the source distribution into three spectroscopic redshift bins of width $0.1$ up to $z=0.3$, plus three photometric 
redshift bins up to $z\simeq0.8$. As described in \S~\ref{sec:pipe}, we then consider two more bins, but that do not correspond to a specific redshift: one for all the radio galaxies in the same sky area covered by SkyMapper/SDSS which are not cross-identified (case \textit{ii} of the ``effective surveys'') and another for all other radio galaxies in patches of the sky not covered by the redshift surveys (case \textit{iii}).

\begin{figure}
\centering
\includegraphics[width=0.5\textwidth]{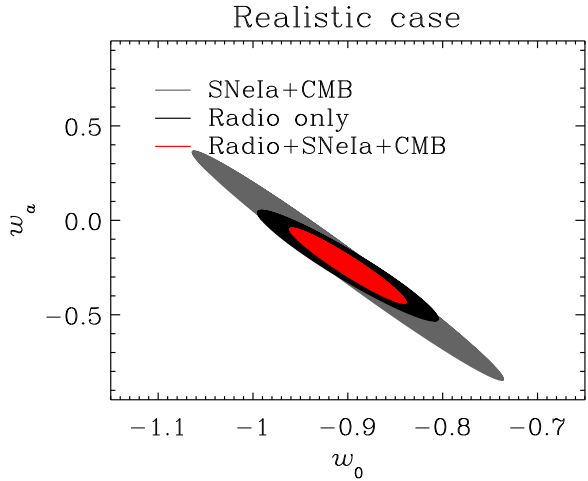}
\caption{Marginal $1$-$\sigma$ contour in the $(w_0,\,w_a)$-plane of the DE model for our realistic binning strategy. The black ellipse refers to the Fisher matrix obtained by using only EMU and WODAN galaxies and their cross-identifications with SkyMapper and SDSS, whilst the red ellipse includes Planck and Union2 priors; for comparison, the grey ellipse shows the contour for Planck and Union2 alone.}\label{fig:ellipses-TOT}
\end{figure}

Fig.~\ref{fig:ellipses-TOT} presents the $1$-$\sigma$ confidence region in the $(w_0,\,w_a)$-plane as obtained by using the radio sources (black ellipse) and when we also add Planck and Union2 (red ellipse); for comparison, we also show the ellipse for Planck and Union2 alone (grey ellipse). The marginal errors $\sigma_{\varphi_i}$ on the fiducial values $\mu_{\varphi_i}$ of the parameters $\varphi_i$, as well as the correlations between the parameters and $w_0$ or $w_a$, are shown in Table~\ref{tab:DE-TOT}. Again, the numbers in brackets refer to results including the full Fisher matrices for Planck and Union2.
\begin{table*}
\caption{The same as Table~\ref{tab:DE} for the realistic binning strategy.}\label{tab:DE-TOT}
\centering
\begin{tabular}{ccrlrlrl}
\hline
$\varphi_i$ & $\mu_{\varphi_i}$ & \multicolumn{2}{c}{$\sigma_{\varphi_i}$} & \multicolumn{2}{c}{$r_{\varphi_i-w_0}$} & \multicolumn{2}{c}{$r_{\varphi_i-w_a}$} \\
\hline
$\om$ & $ 0.28 $ & $ 0.15 $ & $ (0.00056) $ & $ -0.76 $ & $ (-0.46) $ & $ -0.63 $ & $ (0.20) $ \\
$\ob$ & $ 0.045 $ & $ 0.0067 $ & $ (0.0011) $ & $ 0.72 $ & $ (-0.34) $ & $ 0.61 $ & $ (0.26) $ \\
$w_0$ & $ -0.9 $ & $ 0.063 $ & $ (0.041) $ & $ 1 $ & $ (1) $ & $ 0.96 $ & $ (-0.95) $ \\
$w_a$ & $ -0.24 $ & $ 0.195 $ & $ (0.13) $ & $ 0.96 $ & $ (-0.95) $ & $ 1 $ & $ (1) $ \\
$h$ & $ 0.7 $ & $ 0.019 $ & $ (0.00014) $ & $ 0.16 $ & $ (0.345) $ & $ 0.12 $ & $ (-0.26) $ \\
$n_s$ & $ 0.96 $ & $ 0.011 $ & $ (0.00295) $ & $ -0.83 $ & $ (0.37) $ & $ -0.72 $ & $ (-0.17) $ \\
\hline
\end{tabular}
\end{table*}

As expected, these results are not as tightly constraining as those obtained in the idealistic case of a perfect knowledge of all the source redshifts, but are a substantial improvement over the no-redshift case. Indeed, we obtain $\sigma_{w_0}=0.063$ and $\sigma_{w_a}=0.195$, which become $\sigma_{w_0}=0.041$ and $\sigma_{w_a}=0.13$ when combined with Planck and Union2. The correlation coefficients again help to obtain a deeper understanding of these results. As in the idealistic case, the tightness of the constraints on the parameters is fortified by the modest degeneracies yielded by our method. For this purpose, the inclusion of CMB and SNeIa priors is often crucial.

\begin{figure*}
\centering
\includegraphics[width=\textwidth,trim=0mm 50mm 0mm 0mm,clip]{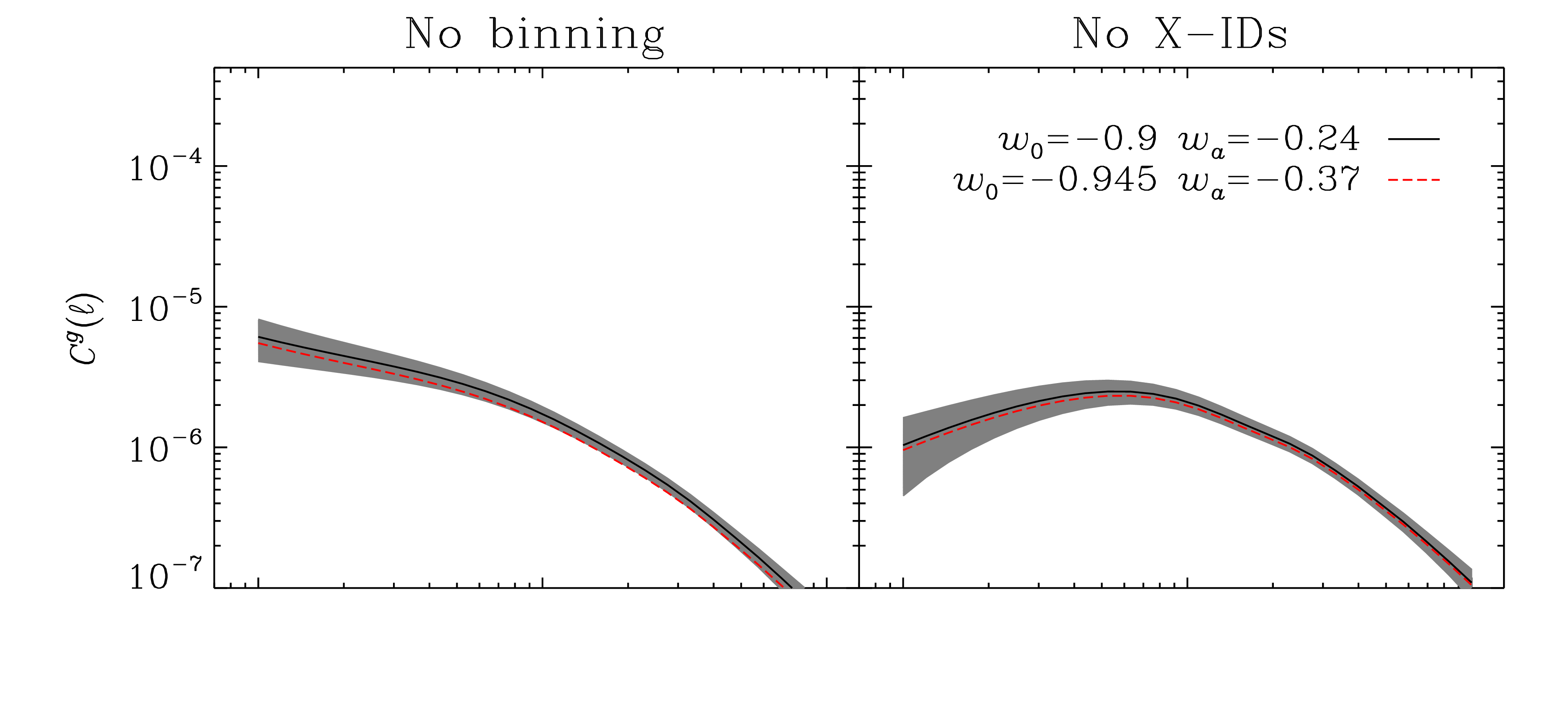}\\
\includegraphics[width=\textwidth]{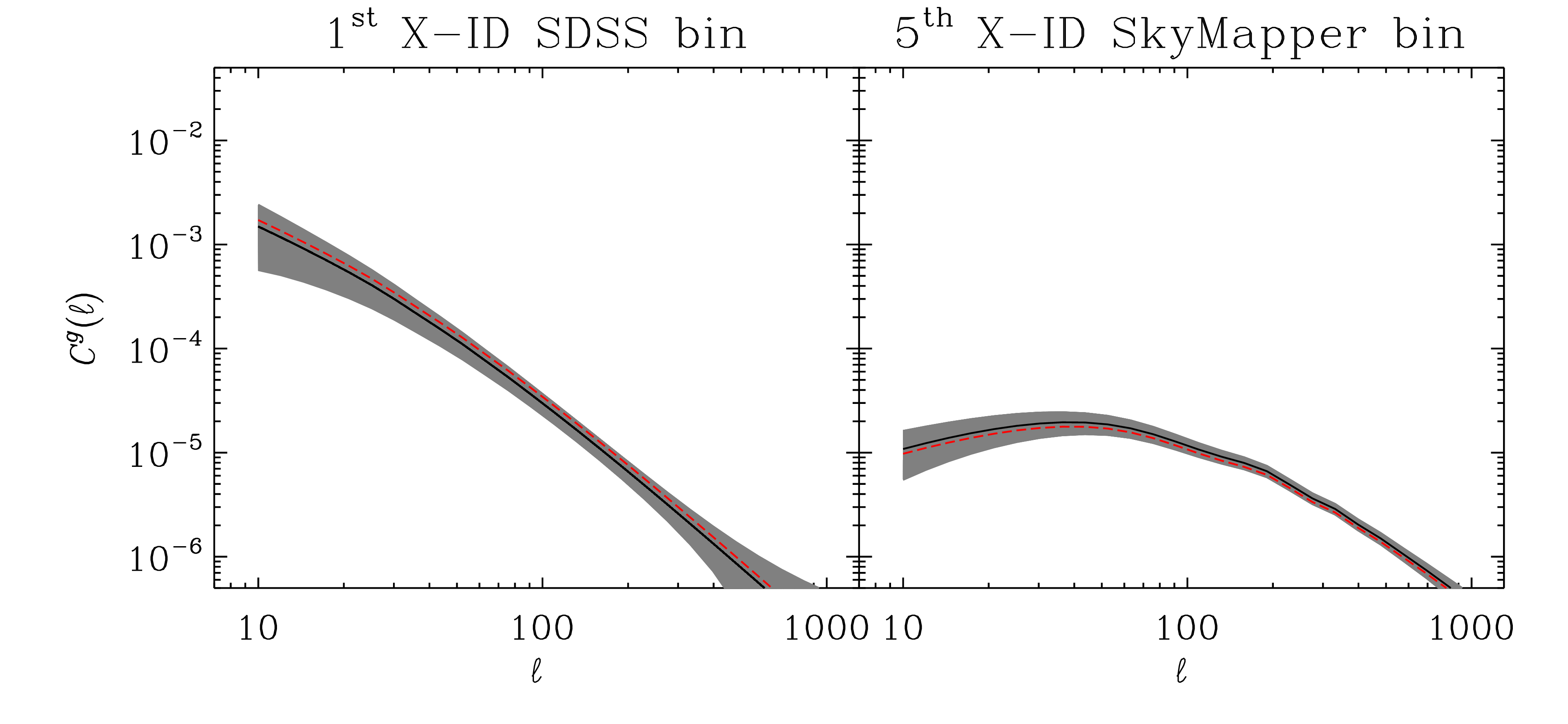}
\caption{Comparison between the angular power spectra for the fiducial DE model (solid, black with errorbars) and a DE cosmology with $\{w_0,\,w_a\}$ values $3\sigma$'s away (dashed, red). \textit{Top-right panel:} EMU+WODAN sources without any redshift information. \textit{Top-left panel:} EMU+WODAN sources not cross-identified by SkyMapper or SDSS. \textit{Bottom-left panel:} first spectro-$z$ bin of the WODAN sources cross-identified by SDSS. \textit{Bottom-right panel:} fifth (and last) photo-$z$ bin of the EMU sources cross-identified by SkyMapper.}\label{fig:comparison}
\end{figure*}

For instance, in Fig.~\ref{fig:comparison} we show, for some of the cases scrutinised, the power spectra $C^g(\ell)$ (solid, black) with the corresponding error bars given by
\begin{equation}
\Delta C^g(\ell)=\sqrt{\frac{2}{(2\ell+1)f_\mathrm{sky}}}\left[C^g(\ell)+\frac{1}{N_g}\right],
\end{equation}
in comparison with a DE model whose $\{w_0,\,w_a\}$ values are $3\sigma$ away according to the errors quoted in Table~\ref{tab:DE-TOT} (dashed, red). Specifically, clock-wise starting from the top-left panel: the power spectrum for the EMU+WODAN sources without redshift information; that for the EMU+WODAN sources not cross-identified by either SkyMapper or SDSS; the $C^g(\ell)$ of the fifth, and last, photometric bin of the sources cross-identified by SkyMapper; and that of the first spectroscopic bin of the radio galaxies cross-identified by SDSS.

We can quantify how much our approach permits a discrimination between DE and \lcdm\ by determining the $\fom$s. In our realistic case, the constraints on $w_0$ and $w_a$ yield $\fom=382$ or $225$ for the radio sources with or without CMB and SNeIa, respectively. All these major results, together with the marginal errors on $w_a$ and on the maximally-constrained value of $w_\mathrm{DE}(z)$---$w_p$ of eq. \eqref{eq:w_p}---are presented in Table~\ref{tab:FoM}.
\begin{table*}
\caption{Summary of the errors on $w_0$, $w_p$ and $w_a$, and the $\fom$s for the unbinned, idealistic and realistic cases without (or with) Planck and Union2.}\label{tab:FoM}
\centering
\begin{tabular}{crlrlrl}
\hline
& \multicolumn{2}{c}{No binning} & \multicolumn{2}{c}{Idealistic case} & \multicolumn{2}{c}{Realistic case} \\
\hline
$ \sigma_{w_0} $ & $ 2.47 $ & $ (0.11) $ & $ 0.022 $ & $ (0.012) $ & $ 0.063 $ & $ (0.041) $ \\
$ \sigma_{w_p} $ & $ 0.29 $ & $ (0.018) $ & $ 0.0090 $ & $ (0.0057) $ & $ 0.020 $ & $ (0.013) $ \\
$ \sigma_{w_a} $ & $ 9.60 $ & $ (0.40) $ & $ 0.058 $ & $ (0.033) $ & $ 0.195 $ & $ (0.13) $ \\
$ \fom $ & $ 0.40 $ & $ (140) $ & $ 1909 $ & $ (5378) $ & $ 260 $ & $ (603) $ \\
\hline
\end{tabular}
\end{table*}

Finally, we want to understand how the assumptions made upon the redshift evolution of the bias influence our results. To this aim (see \S~\ref{ssec:bias}) we make use of the additional nuisance parameter $B_b$, as in Eq.~\eqref{eq:lin-bias}, thus introducing a further redshift dependence of the bias. We again perform the same Fisher analysis as before, where the set of nuisance parameter we want to marginalise over is now $\bpsi=\{\sigma_8,\,A_b,\,B_b\}$. As expected, the presence of one more free parameter leads to weaker constraining power. As a result, the $w_0$-$w_a$ ellipse is broader than in the realistic case. Nevertheless, we want to demonstrate that, even in this pessimistic case, our approach gives interesting results.

\begin{figure}
\centering
\includegraphics[width=0.5\textwidth]{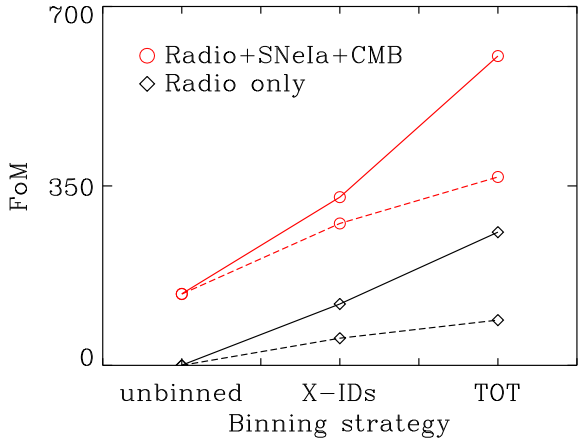}
\caption{$\fom$s with or without the inclusion of Planck and Union2 (red circles or black diamonds, respectively) versus the binning strategy adopted: ``unbinned'' pertains to the case with no redshift information, ``X-IDs'' to the EMU/WODAN sources cross-identified by SkyMapper/SDSS and ``TOT'' to the inclusion of all the effective surveys. Solid lines refer to the main case, whilst dashed lines to the case with an additional linear parameterisation for the bias.}\label{fig:FoM}
\end{figure}

Therefore, to conclude, Fig.~\ref{fig:FoM} summarises the primary results of this work. It depicts the $\fom$ versus the binning strategy adopted, thus showing the impact of our proposed pipeline. The labels of the horizontal axis refer to the unbinned case (unbinned), the case with cross-identified sources alone (X-IDs)  and the case resulting from the combination of all the effective surveys (TOT)---\textit{i)}+\textit{ii)}+\textit{iii)} as explained in \S~\ref{sec:pipe}. Black diamonds refer to the use of only EMU and WODAN galaxies and their cross-identifications with SkyMapper and SDSS, whilst red circles point to the results obtained by the inclusion of the Fisher matrices for Planck and Union2. Solid lines refer to the main result for the realistic case, as highlighted in the previous section; dashed lines instead show what happens in the case just described of a two-parameter bias. It is clear that the inclusion of cross-identifications and, furthermore, the high-$z$ radio tail still yields a 
significant enhancement.

\section{Summary and Conclusions}\label{sec:conclusions}
In this paper, we have explored the impact that redshift information could have on the cosmological potential of the forthcoming generation of large-scale, radio continuum surveys. Specifically, we have investigated how the possibility of binning the redshift distribution of the radio sources would improve the constraining power of these surveys. For this purpose, as a fiducial model we adopted a dynamical DE model whose fiducial values have been fitted by \citet{Zhao:2009ti} as $\{w_0,\,w_a\}=\{-0.9,\,-0.24\}$. We have focused on two forthcoming wide-field radio surveys: EMU and WODAN. Thanks to their sensitivity and the specifics outlined in Table~\ref{tab:EMUWODAN}, they could be combined thus yielding a full-sky survey. This is extremely important, since it leads to larger sample size and allows us to lessen the impact of cosmic variance.

We have performed a Fisher matrix analysis to forecast how the combination of EMU and WODAN will be able to constrain the cosmological parameters of the DE model, particularly focusing on its extra parameters, $w_0$ and $w_a$, and using the angular power spectrum of radio sources as the cosmological probe. We have examined the impact of including redshifts for sources cross-identified with optical redshift surveys. We have described results when there is no knowledge of redshifts, summarised in Table~\ref{tab:DE}; in this case the DE equation of state is poorly determined. We have also presented the most idealistic case where all the EMU and WODAN redshifts are known, in which case we have unsurprisingly obtained extremely tight constraints on all the model parameters, with marginal errors on the DE equation-of-state parameters $\sigma_{w_0}=0.012$ and $\sigma_{w_a}=0.033$, when combined with Planck and Union2.

Next, we moved to a more realistic scenario. We have included forthcoming optical surveys as cross-identifiers for the EMU/WODAN sources; specifically, we have used SkyMapper for the southern sky, while SDSS provides redshifts in the north. It is clear that not all of the radio galaxies that will be detected by EMU and WODAN will be cross-identified by other surveys, as these radio surveys cover the whole sky and reach very high redshift. To take this into account, we have developed a model of the future practical analysis, using three ``effective surveys'' as described in \S~\ref{sec:pipe}. For these sub-surveys, the redshift distribution of sources will be those of: \textit{i)} all the EMU/WODAN galaxies cross-identified by SkyMapper/SDSS, suitably redshift binned; \textit{ii)} all the EMU/WODAN galaxies which are in the same patch of sky as SkyMapper/SDSS but which are not cross-matched; and \textit{iii)} an EMU/WODAN survey probing only the sky left uncovered by SkyMapper/SDSS. With this approach we have 
obtained a FoM of $603$ and $260$ for the radio sources with and without CMB and SNeIa priors, respectively. These results are very competitive when compared with other surveys that will also be available in the near future. The improvement obtained by this cosmological analysis with radio surveys stems from the fact that by cross-correlating with shallow ``redshift surveys'' we can make use of the long high redshift tail of the non-identified radio galaxies to provide an extra handle on the evolution of DE. This effect can be clearly seen in Fig.~\ref{fig:FoM}, where the jump from the second to the third point is only due to the extra information from the radio-surveys.

We want to emphasise that we have accounted for the source bias amplitude $A_b$ by using a nuisance parameter which was allowed to vary freely. However, reality might be more complicated with effects on the bias such as scale dependence and a change in the relative amplitudes of the bias for each finer redshift bin. We leave a deeper analysis for future work. It is worth noting that the radio source bias could be constrained as a function of redshift by deep cross-correlation studies with optical data over smaller areas, but any remaining uncertainty requiring more bias parameters will worsen the $\fom$s. To simply test whether our approach is robust even in the presence of a more complex bias, we have also studied a slightly more complex scenario, in which we have added a linear correction to the bias amplitude. This means that we deal with one more nuisance parameter. Nevertheless, Fig.~\ref{fig:FoM} shows that the main results of this work still hold, though the constraints get broader, as expected. On 
the other hand, if we parameterised the functional form of the bias not only by its amplitude but also taking its shape into account, we should be able to lift the strong degeneracy present here between $A_b$ and the matter power spectrum normalisation $\sigma_8$.

We have also not accounted for possible contaminants to the power spectrum that could affect our analysis, since the objective of this paper is to concentrate on the impact of redshift information for cosmological applications. In particular, double sources could affect the angular power spectrum at smaller angular scales ($<0^\circ.1$), requiring some cleaning technique \citep[see e.g.][for the case with NVSS data]{Overzier:2003kg}. Note, however, that for the $10\sigma$-$10\,\mu\mathrm{Jy}$ flux cut we are considering here, these double sources should be much less abundant and much less dominant than for the NVSS case, so that the contamination might be less severe on these sub-degree scales; moreover, we are cutting off our analysis at $\ell_\mathrm{max}=1000$.

Besides, instrumental effects may also alter the clustering signal. For instance, \citet*{Blake:2003dv} have pointed out that, for angular separations around $\sim0^\circ.2$, there is an unexpected deficit in NVSS data, which is very likely an instrumental effect due to the shortest baseline used in the observations. Similarly, \citet{Blake:2001bg} have shown the risk in not properly considering and modelling gradients in the surface density. Indeed, such effects may significantly influence the imprint of the large-scale structure. This is because, if we tried to fit the data with a theoretical angular power spectrum which does not account for these systematics, we would reconstruct the systematic density gradients as well as the fluctuations due to clustering. However, such problems should arise with flux cuts $\gtrsim10\,\mathrm{mJy}$ \citep*[see also][]{Blake:2003dv,Blake:2004dg}, thus leaving our analysis free of this problem. Nevertheless, we plan to return to the specific issue of 
contamination and cleaning of the cosmological signal in a future analysis.

As a final remark, we note that the optical surveys we have used for the cross-identifications should be available at approximately the same time as EMU and WODAN, around $2015$. Therefore, we can see that future large-scale radio surveys can not only provide a different observational wavelength window for constraining the cosmological model, but also provide competitive DE constraints. Related to this, we aim to exploit such a successful method in a forthcoming paper to also test modified gravity models \citep{Camera:radio-mg}.

\subsection*{Acknowledgments}
We thank the referee for a careful reading of our manuscript and very insightful comments. SC and MGS acknowledge support from FCT-Portugal under grant PTDC/FIS/100170/2008. SC's work is funded by FCT-Portugal under Post-Doctoral Grant SFRH/BPD/80274/2011. Part of the research described in this paper was carried out at the Jet Propulsion Laboratory, California Institute of Technology, under a contract with the National Aeronautics and Space Administration. The authors would also like to thank Gong-Bo Zhao for his friendly and gratuitous contribution.

\bibliographystyle{mn2e}
\bibliography{/home/stefano/Documents/LaTeX/Bibliography}

\bsp

\label{lastpage}

\end{document}